\documentclass[a4,11pt]{cip-submit}  
\usepackage{afterpage}

\usepackage{mathptmx}
\usepackage{hyperref}
\hypersetup{
	colorlinks=false,
	pdfborder={0 0 0}
}
\def\Mr{\uppercase}

\usepackage{float}
\usepackage{wasysym}
\usepackage{graphicx,wrapfig,tikz}
\usepackage{caption}
\usepackage{subcaption}
\usepackage{amsmath,amsxtra,amssymb,latexsym,amscd,color,bm}
\usepackage[mathscr]{eucal}
\usepackage{epstopdf}
\usepackage{makecell}
\usepackage{multirow}
\usepackage{ulem}
\usepackage{xcolor}
\usepackage{nicefrac}
\usepackage{array}
\usepackage{cite}

\def\vsm{\vskip0.1cm}
\def\doi#1{\href{http://dx.doi.org/#1}{doi:#1}}

\def\titles#1{\title{\large\bf\noindent #1}}
\def\authors#1{\author{\begin{flushleft}{#1}\end{flushleft}}}
\def\authord#1#2{\indent\Mr{#1}\\
	\textit{\indent#2}\vsm}

\def\email#1{\bigskip\href{mailto:#1}{\textit{E-mail:}~{#1}}\\[3mm]}

\def\received#1{\vsm\textit{\indent Received #1}}
\def\accepted#1{\vsm\textit{Accepted for publication~#1}}

\def\Keywords#1{\\[.2cm] Keywords:~{#1}.}

\def\and{$\text{\tiny AND }$}

\def\Classification#1{\\[.2cm] Classification numbers:~{#1}.} 
\newcommand{\kms}{km s$^{-1}$}
\newcommand{\dego}{$^\circ$}

\begin{document}
	\Year{2020}
	\Page{1}\Endpage{12}
	\titles{Observation of high Doppler velocity wings in the nascent wind of R Doradus}
	\authors{	\authord{D.T. Hoai, P.T. Nhung, P. Tuan-Anh, P. Darriulat, P.N. Diep, N.T. Phuong and T.T. Thai} {Vietnam National Space Center (VNSC), Vietnam Academy of Science and Technology (VAST), 18 Hoang Quoc Viet, Cau Giay, Ha Noi, Vietnam\\
          }
	        \email{dthoai@vnsc.org.vn}
		\received{\today}
		\accepted{DD MM YYYY}
	}
	\maketitle
	\markboth{D. T. Hoai}{High Doppler velocity wings in the nascent wind of R Doradus}

\begin{abstract}
We study the morpho-kinematics in the nascent wind of AGB star R Doradus in the light of high Doppler velocity wings observed in the spectral lines of several species. We probe distances from the star between $\sim$10 and $\sim$100 au using ALMA observations of the emission of five different molecular lines. High Doppler velocity enhancements of the line emission are observed in the vicinity of the line of sight crossing the star, reminiscent of those recently interpreted as gas streams in the nascent wind of a similar AGB star, EP Aqr. They are present in both blue-shifted and red-shifted hemispheres but are not exactly back-to-back. They are accelerated at a typical rate of 0.7 \kms au$^{-1}$ up to some 20 \kms. Important differences are observed between the emissions of different molecules. We exclude an effect of improper continuum subtraction. However, in contrast to EP Aqr, the line of sight plays no particular role in the R Dor morpho-kinematics, shedding doubt on the validity of a gas stream interpretation.We discuss possible interpretations in terms of stellar pulsations or of rotation of the gas in the environment of the star. We conclude that, in the state of current knowledge, no fully convincing picture of the physics governing the production of such high velocities, typically twice as large as the terminal velocity, can be reliably drawn. New high resolution analyses of observations of the nascent wind of oxygen-rich AGB stars are needed to clarify the issue. 
\Classification{97.60-s}
\end{abstract} 



\section{INTRODUCTION}\label{sec1}

The occasional presence of large Doppler velocity wings in some line emission spectra of oxygen-rich evolved stars has been known for some time \cite{Cernicharo1997}. They are particularly visible in SiO line profiles and may typically reach twice the terminal wind velocity. Single dish millimetre observations analysed by Winters et al. (2003)\cite{Winters2003} and more recently by de Vincente et al. (2016)\cite{deVicente2016} have suggested that the emission occurs close to the star, where SiO grains have not yet fully formed, and is somehow related to star pulsations.

In the past few years, the availability of high angular resolution ALMA observations of the nascent winds of two similar AGB stars, EP Aqr and R Dor, has shed new light on this issue.

The presence of high Doppler velocity components in the $^{28}$SiO($\nu=0, J=5-4$) line emission of the nascent wind of EP Aqr was first noted by Homan et al. (2018a)\cite{Homan2018a}  in Section 4.4 of their article. The morpho-kinematics of the circumstellar envelope of EP Aqr is known to display axi-symmetry about an axis close to the line of sight \cite{Hoai2019}. The study of SO$_2$ line emission very close to the star gives evidence for rotation about the same axis \cite{Homan2018a, TuanAnh2019}. Inhomogeneity has been revealed in the equatorial plane, close to the plane of the sky, in the form of a spiral of intensity \cite{Homan2018a} and of concentric rings of radial velocity \cite{Hoai2019}. The spiral of intensity was tentatively interpreted by Homan et al. (2018a)\cite{Homan2018a} as evidence for the presence of a companion. In such a context it was therefore natural for Homan et al. (2018a)\cite{Homan2018a} to consider the possibility that the complex dynamics in the wind-companion interaction zone not only accelerate a large portion of the outflow material along the equatorial plane, but also along the polar axis, causing some of this material to have increased velocities along the line of sight.

Tuan-Anh et al. (2019)\cite{TuanAnh2019} have then performed a detailed study of the morpho-kinematics of the high velocity wings of the $^{28}$SiO($\nu=0, J=5-4$) line emission and have shown that they are also present, but much weaker, in the $^{12}$CO(2-1) line emission. Their absence from the SO$_2$ line emission, at very short distances from the star, disfavours too sudden an acceleration and, in particular, an interpretation in terms of star pulsation. Tuan-Anh et al. (2019)\cite{TuanAnh2019} propose instead a picture of two narrow polar streams of gas, referred to as jets, being launched from less than 25 au away from the star, building up between $\sim$20 au and $\sim$100 au to a velocity of $\sim$20 \kms\ and fading away at larger distances. 

Both Homan et al. (2018a)\cite{Homan2018a} and Tuan-Anh et al. (2019)\cite{TuanAnh2019} insist on the complexity of the morpho-kinematics of the nascent wind of EP Aqr when observed with the high angular resolution offered by ALMA and on the difficulty to draw a convincing and reliable picture of the physics at stake.

Decin et al. (2018)\cite{Decin2018} were first to discuss the presence of high velocity wings in some line emission spectra of the nascent wind of R Dor,  far above the canonical terminal wind velocities although hints of their existence had been mentioned earlier (e.g. Justtanont et al. 2012\cite{Justtanont2012}).  They quote a maximal Doppler velocity of 23 \kms\ reached in the line emission of SiO($\nu=0, J=8-7$) (note that the velocity scale in their Figure 8 is a factor 2 too large, we thank Pr Leen Decin for clarification on this point). They show in their Figure 10 typical values of the end point velocity of different line spectra reaching some $\sim$15 \kms. On the basis of a model proposed by Nowotny et al. (2010)\cite{Nowotny2010} they argue that star pulsations cannot significantly contribute to the generation of such high velocities. They conclude that the origin of the large velocities is a genuine physical mechanism not linked to thermal motions of the gas or pulsation behaviour of the atmospheric layers and that their impact on the mass-loss rate cannot be underestimated.

They privilege a scenario developed by Homan et al. (2018b)\cite{Homan2018b} suggesting the presence of a rotating disc, seen nearly edge-on, in the close neighbourhood of the star. It covers radial distances between 6 and 25 au and the Doppler velocity is maximal on the inner rim where it reaches 12 \kms. The relation between this disc and the high Doppler velocity wings is the suspected presence of an evaporating companion planet providing the necessary angular momentum to the disc rotation. Evidence for the possible existence of such a companion rests on the observation of high Doppler velocity emission in the blue-shifted hemisphere near the middle of the south-eastern quadrant. This enhanced emission is referred to as the ``blue blob'' by both Decin et al. (2018)\cite{Decin2018} and Homan et al. (2018b)\cite{Homan2018b}. They discuss critically the properties of the ``blue blob'' and carefully conclude that follow-up high-resolution observations are needed to test their claims, and to deeper investigate the true nature of both the disk-like signal and the blue blob.

Finally, Vlemmings et al. (2018)\cite{Vlemmings2018} analyse ALMA observations made two years later with a four times better angular resolution than Decin et al. (2018)\cite{Decin2018}, $\sim$35 mas instead of $\sim$150 mas. From continuum emission they resolve the stellar radio photosphere as a circle of 31 mas radius. The analysis covers the compact emission of two lines: SiO($\nu=3, J=5-4$) and SO$_2$($J_{Ka,Kc}=16_{3,13}-16_{2,14}$) and the absorption of a third, $^{29}$SiO($\nu=1, J=5-4$). The line emissions are shown to be properly described by solid body rotation of expanding shells having average radii of 36 mas (2.2 au) and 66 mas (4.0 au) for SiO($\nu=3, J=5-4$) and SO$_2$($J_{Ka,Kc}=16_{3,13}-16_{2,14}$) respectively. The authors claim that their result contradicts the interpretation made by Homan et al. of a nearly edge-on rotating disc. Their model implies a positive radial velocity gradient, at variance with the disc model of Homan et al. (2018b)\cite{Homan2018b} that implies a negative radial velocity gradient. However, following Homan et al., they suggest that the most likely cause of the observed rotation is the presence of a companion. They show a position-velocity diagram illustrating the region of the ``blue blob'' discovered earlier by Decin et al. (2018)\cite{Decin2018}. This diagram is using $^{29}$SiO($\nu=0, J=5-4$) observations about which no detail is given. The Doppler velocity reaches some 15 \kms. The authors discuss it in the framework of their solid body rotation model and recognize that the origin of the fast rotating feature out to $>10R_*$ remains unclear and that this feature could be unrelated to the rotation and represent a seemingly one-sided ejection of material. Apart from the discussion of this feature, the authors do not mention the possible existence of high Doppler velocity wings  but simply quote line widths of $\sim5-10$ \kms\ for the SO$_2$ line, $\sim$10 \kms\ for the SiO $\nu=3$ line, and $\sim15-20$ \kms\ for the $^{29}$SiO $\nu=1$ line. A fortiori, they do not discuss possible causes for their existence, such as the effect of stellar pulsations.

The lack of a convincing physical picture of the mechanism governing the production of high Doppler velocity wings in the nascent winds of EP Aqr and R Dor, together with the absence of a detailed and dedicated analysis of their morpho-kinematics in the case of R Dor, have motivated the present work.

EP Aqr and R Dor are both semi-regular variables of the SRb type and belong to similar spectral classes, M8IIIvar and M8IIIe. They have similar initial masses, between 1 and 2 M$_\odot$, similar mass loss rates, $\sim$1.6 10$^{-7}$ M$_\odot$ yr$^{-1}$ \cite{Hoai2019, Maercker2016} and similar temperatures, within 100 K from 3150 K \cite{Dumm1998}. They are both close to the Sun, at distances of $\sim$119 pc \cite{vanLeeuwen2007,Gaia2018} and $\sim$59 pc \cite{Knapp2003}, respectively. Both display no technetium in their spectrum \cite{Lebzelter1999} and have the same $^{12}$CO/$^{13}$CO abundance ratio of $\sim$10 \cite{TuanAnh2019, Ramstedt2014}. They differ by their pulsation period, 55 days for EP Aqr \cite{Lebzelter2002} and a dual period of 175 and 332 days for R Dor \cite{Bedding1998} but their infrared emissions above black body between 1 and 40 $\mu$m wavelength are similar \cite{Heras2005}, the main difference being a relative enhancement of silicates and depression of aluminium oxide in the EP Aqr dust.

The morpho-kinematics of the circumstellar envelope of EP Aqr has been measured up to some 1000 au from the star \cite{Homan2018a, Nhung2019a, Hoai2019, TuanAnh2019}. It is dominated above 200 au by an axi-symmetric radial wind having reached a terminal velocity of $\sim 2+9\sin^2\alpha$ \kms, $\alpha$ being the stellar latitude and the polar axis making an angle of $\sim$10\dego\ with the line of sight. The circumstellar envelope of R Dor has been probed by ALMA with a resolution of $\sim$4 au up to some 60 au from the star \cite{Decin2018, Homan2018b, Vlemmings2018} and the dust has been observed at the VLT with a resolution of 1.2 au \cite{Khouri2016}. In addition, below $\sim$15 au, the analyses of Danilovich et al. (2016)\cite{Danilovich2016}, De Beck \& Olofsson (2018)\cite{DeBeck2018} and Van de Sande et al. (2018)\cite{VandeSande2018} have contributed a considerable amount of detailed information of relevance to the physico-chemistry and dynamics of both dust and gas. At larger distances from the star, an analysis of ALMA observations of SO($J_K=6_5-5_4$) emission \cite{Nhung2019b}, probing distances between 20 and 100 au, gives evidence for the wind to host a radial outflow covering large solid angles and displaying strong inhomogeneity both in direction and radially: the former takes the form of multiple cores and the latter displays a radial dependence suggesting an episode of enhanced mass loss having occurred a century or so ago.

In what follows, we explore a possible interpretation of the high velocity wings present in the nascent wind of R Dor in terms similar to those observed in EP Aqr. We study the large Doppler velocity features displayed by line emissions detected between $\sim$15 and $\sim$50 au from the star \cite{Decin2018}, which suggest similarities with the EP Aqr dynamics. These include excitations of five different molecules: CO, SiO, SO, SO$_2$ and HCN.

\section{OBSERVATIONS AND DATA REDUCTION}\label{sec2}

The data are retrieved from ALMA archives.

The SO data (time on source of 2.7 hours) are from project 2017.1.00824.S observed in December 2017 in band 6 with an average of 45 antennas. They have been used by Nhung et al. (2019b)\cite{Nhung2019b} to study the slow wind using the calibrations and deconvolution provided by the standard ALMA pipeline. We have checked the quality of the data reduction and evidence for the Gaussian distribution of the noise  and for the proper description of the continuum map is given there, together with channel maps. These data have not been used by other authors and the study of Nhung et al. (2019b)\cite{Nhung2019b} does not address the issue of large Doppler velocity components.

All other data (time on source of $\sim$25 minutes) are from project 2013.1.00166.S observed in summer 2015 in band 7 with an average of 39 antennas (Table \ref{tab1}). They have been used by Decin et al. (2018)\cite{Decin2018} who show channel maps but do not perform a detailed analysis of the large Doppler velocity components. These observations, reduced using the results of the standard ALMA pipeline, include datasets associated with significantly different $uv$ coverage, implying different maximal recoverable scales. While this implies important differences in the flux-density measured in the region of the slow wind, we have checked that very similar results are instead obtained in the region of large Doppler velocities explored here at small projected distance from the star.

In the present work we also use SiO line data from the same project reduced by us without subtracting the continuum. They were calibrated from the raw data available in the archive and deconvolved using the same procedure as for the continuum subtracted data.

\begin{table*}
  \centering
  \caption{Line emissions considered in the present work. All are in the vibrational ground state.}
  \label{tab1}
  \vspace{0.5cm}
  
  \begin{tabular}{cccccc}
      \hline
      Line&CO(3-2)&SiO(8-7)& \makecell{SO\\($J_K=6_5$-$5_4$)}&\makecell{SO$_2$\\($13_{4,10}$-$13_{3,11}$)}&HCN(4-3)\\
      \hline
      Frequency (GHz)&345.796&347.331&251.826& 357.165&354.505\\
      \hline
      Beam FWHM (mas$^2$)&180$\times$140&180$\times$130&154$\times$147&160$\times$130&157$\times$145\\
      \hline
      \makecell{Noise \\(mJy\,beam$^{-1}$\,channel$^{-1}$)}& 5.5&4.8&1.1&6.3&8.5\\
      \hline
      \makecell{Channel spacing\\ (\kms)}&0.42& 0.42&0.29&0.41&0.41\\
      \hline
      \makecell{Peak intensity\\(Jy beam$^{-1}$)}&0.74&1.31&0.13&0.17&0.39\\
      \hline
      $E_u/k$ (K)& 33.2&75.0&50.7&123&42.5\\
      \hline
      
  \end{tabular}
\end{table*}

\section{LARGE DOPPLER VELOCITY COMPONENTS}\label{sec3}

We use coordinates rotated to have the ``blue-blob'' detected by Decin et al. (2018)\cite{Decin2018} at a position angle of approximately 180\dego, meaning that the $x$ axis points 35\dego\ north of east and the $y$ axis 35\dego\ west of north; the $z$ axis points away from us, parallel to the line of sight, and the origin of coordinates is taken at the centre of continuum emission. Doppler velocity ($V_z$) spectra are referred to a local standard of rest velocity of 7.0 \kms. 

\subsection{Interpretation in terms of gas streams}\label{sec3.1}

\begin{figure*}
  \centering
   \includegraphics[width=0.9\textwidth]{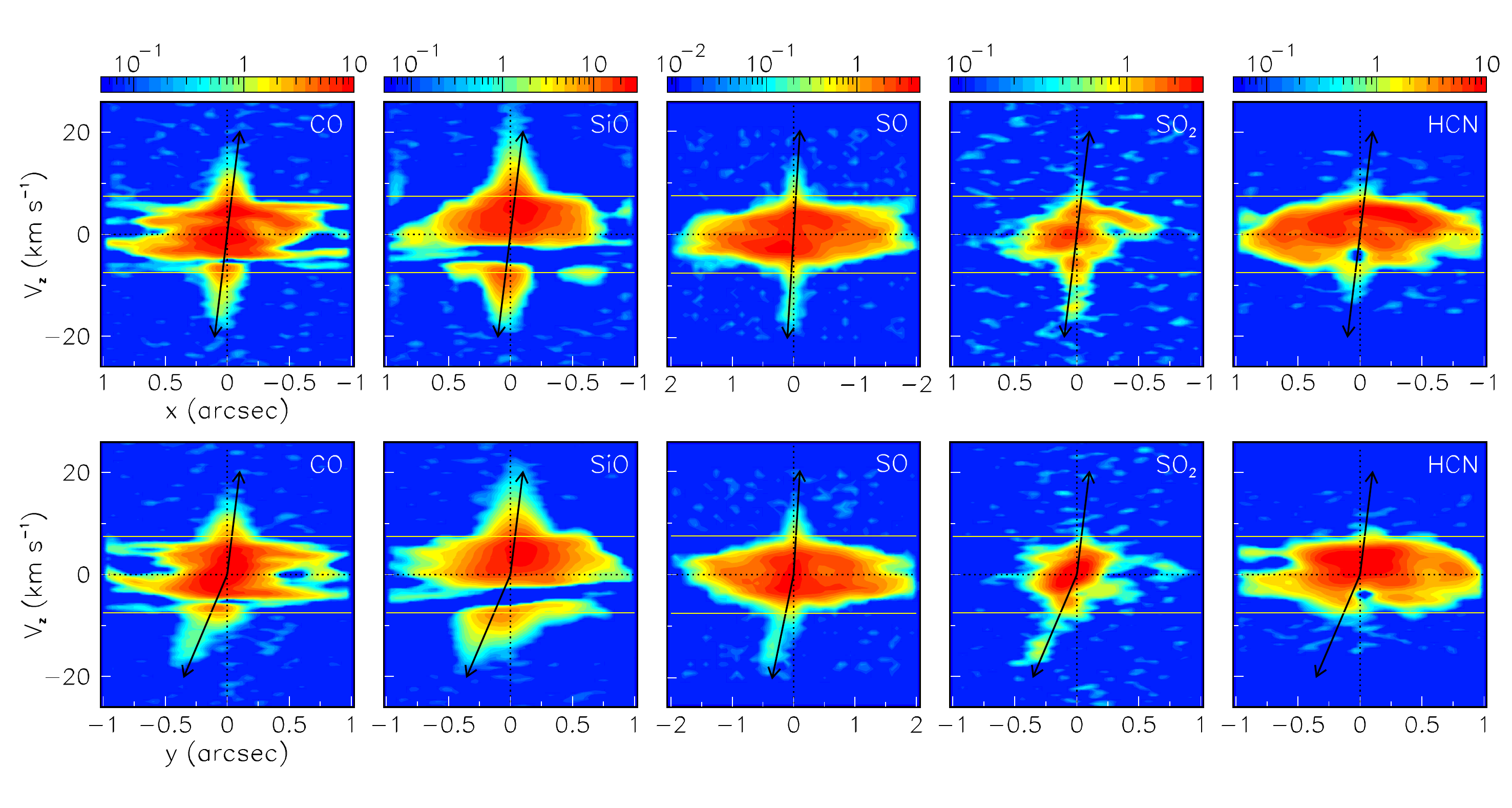}
  \caption{P-V maps in the $V_z$ vs $x$ and vs $y$ planes. The colour scale is in units of Jy arcsec$^{-1}$. Yellow lines show the cuts applied in the definition of the large Doppler velocity components. The SO map extends up to 2 arcsec. The black arrows cover from the origin to $(x,y)$=(0.1,$-$0.35) arcsec and $V_z=-20$ \kms\ in the blue-shifted hemisphere and to $(x,y)$=( $-$0.1,0.1) arcsec and $V_z$=20 \kms\  in the red-shifted hemisphere.}
  \label{fig1}
\end{figure*}

\begin{figure*}
  \centering
   \includegraphics[width=0.95\textwidth,trim=0.cm .5cm 0.cm 0.5cm,clip]{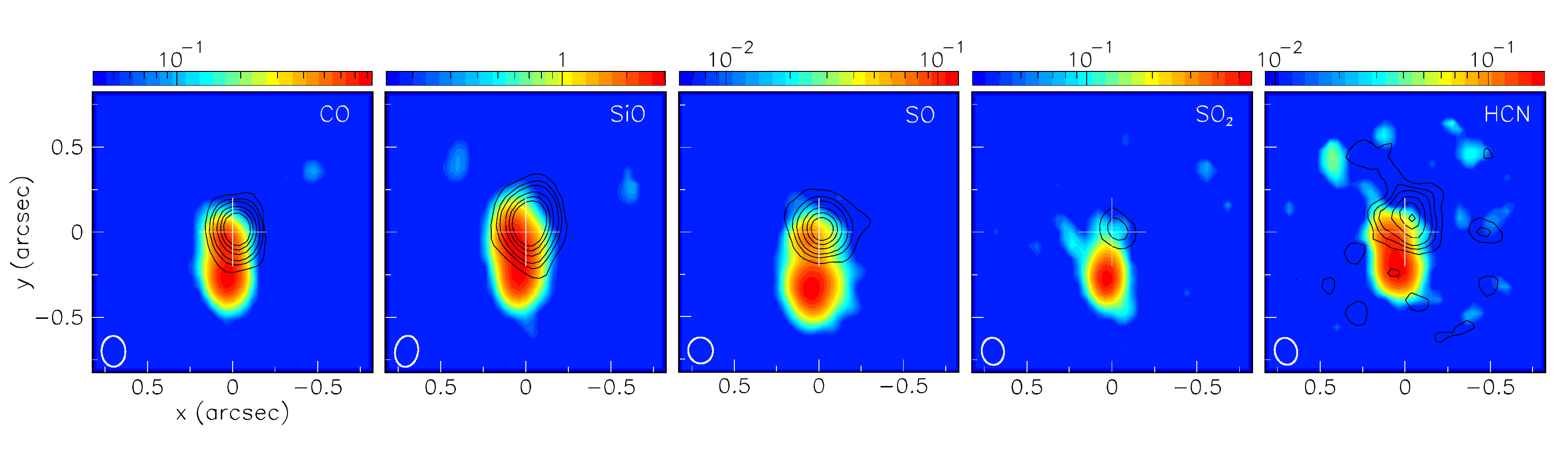}
  \caption{Intensity maps of the high $|V_z|$ components. Contours show the red-shifted stream, the colour maps show the blue-shifted stream. The colour scale is in units of Jy beam$^{-1}$ \kms. The contour levels are at 10\%, 20\%, 30\%, 50\%, 70\% and 90\% of the peak intensity of the blue-shifted stream. The beams are shown in the lower left corners. The white crosses mark the position of the continuum peak.}
  \label{fig2}
\end{figure*}

\begin{figure*}
  \centering
  \includegraphics[height=9cm,trim=1.2cm 0.cm 0.cm 1.cm,clip]{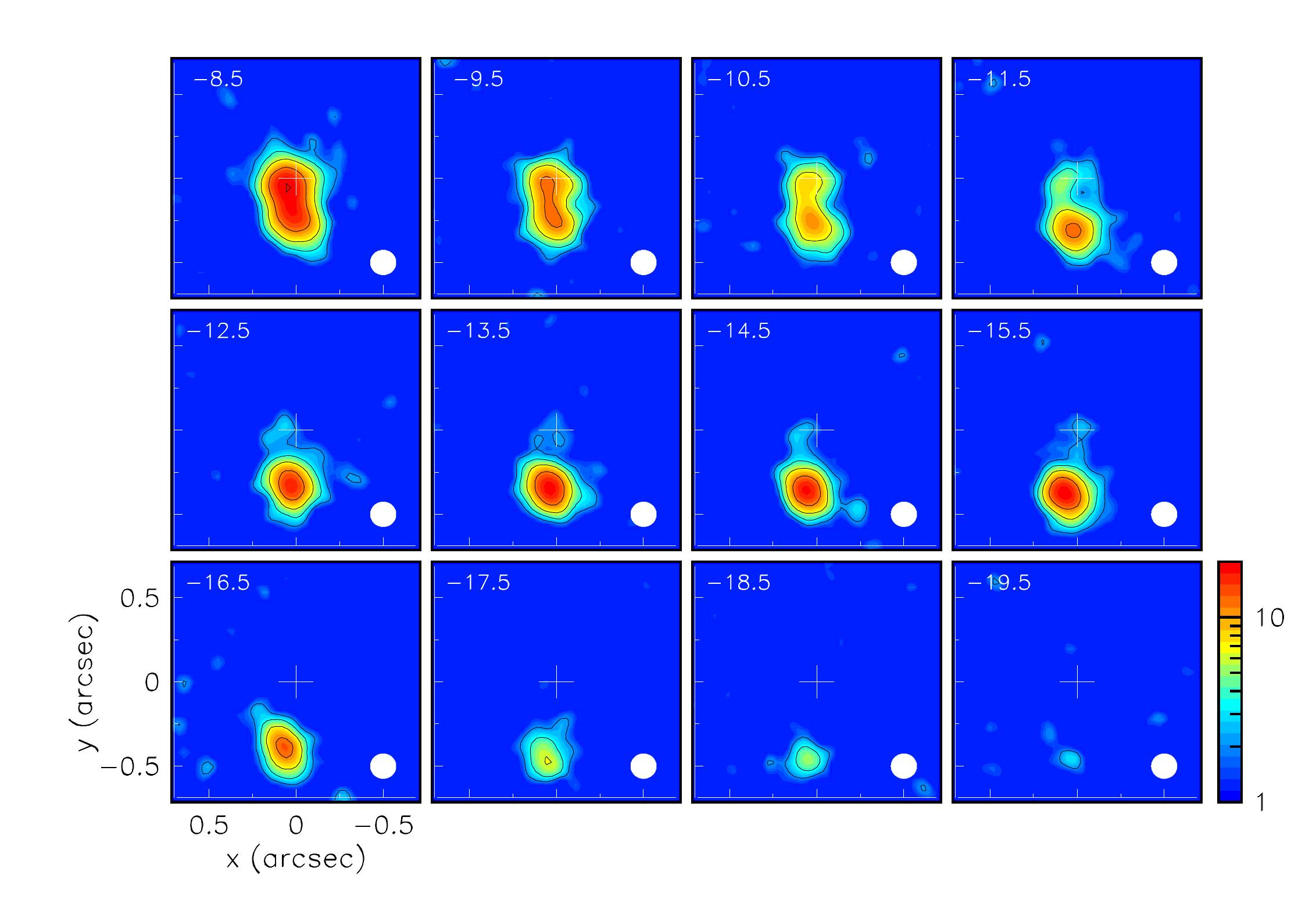}
  \includegraphics[height=9cm,trim=1.2cm 0.cm 0.cm 1.cm,clip]{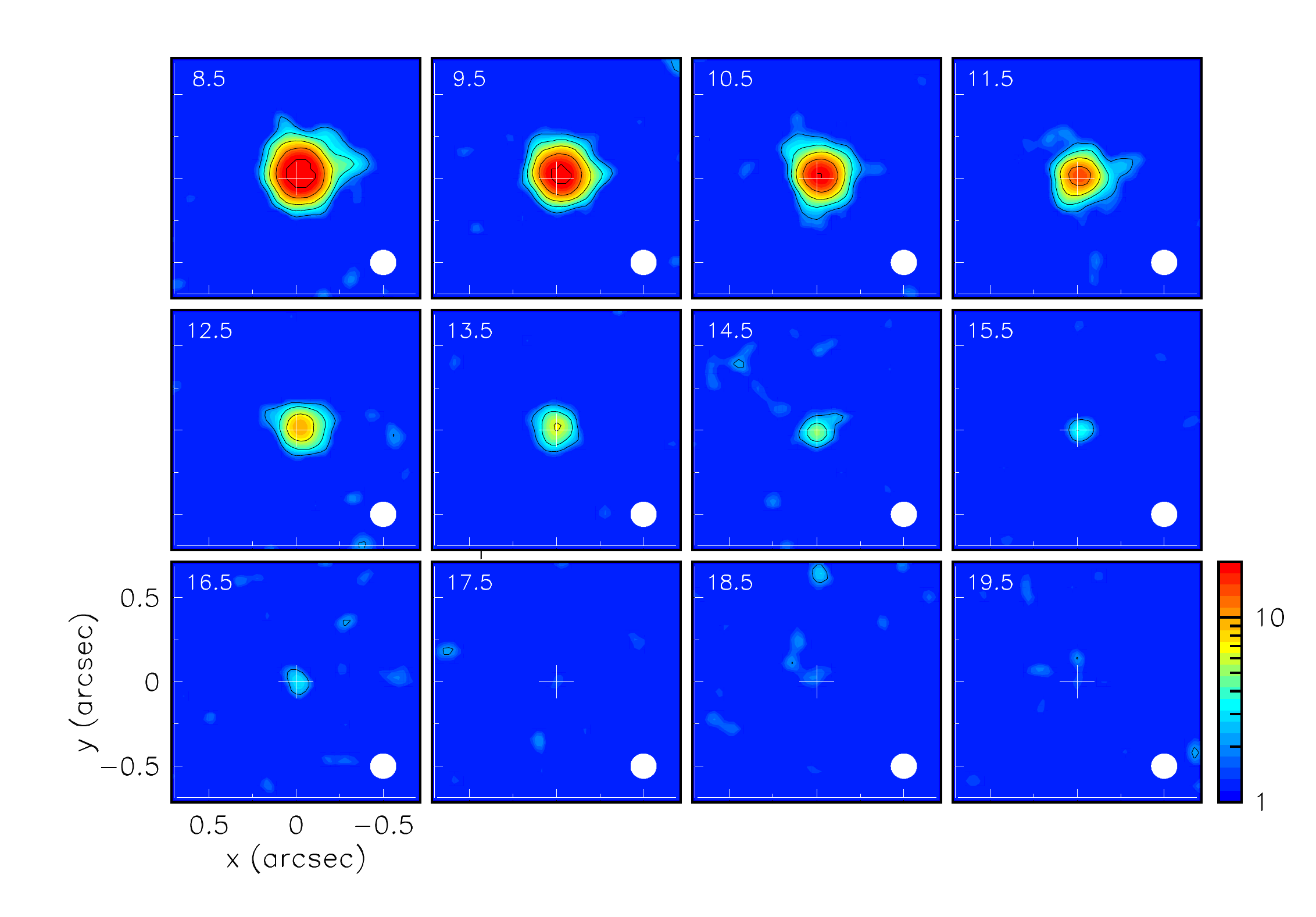}
  \caption{Channel maps of the large Doppler velocity components of the SO line in the blue-shifted (upper panels) and red-shifted (lower panels) hemispheres. For each hemisphere we use 1 \kms\ bins in $|V_z|$, the velocity is indicated in the upper left corner of each panel.  The common colour scale is in units of mJy beam$^{-1}$. The white crosses mark the position of the continuum peak.}
  \label{fig3}
\end{figure*}

Projections of the data-cubes on the ($x, V_z$) and ($y, V_z$) planes, to which we refer as P-V maps, are shown in Figure \ref{fig1}. In contrast with standard P-V diagrams, these are not restricted to narrow slits but are summed over the data-cube, namely integrated over $y$ and $x$ respectively. In all cases the larger values of $|V_z|$ are confined near $x= 0$, very much as was observed in EP Aqr \cite{TuanAnh2019}. We define large Doppler velocity components as having $|V_z|>7.5$ \kms\ in order to separate them from the slower wind. Figure \ref{fig2} displays the maps of their integrated intensity. Together with Figure \ref{fig1}, they show an accelerating stream-like morphology, which was already apparent from the progression of the ``blue-blob'' toward the star at a rate of $\sim$0.7 \kms au$^{-1}$ in the SO$_2$ channel maps displayed in Figure B1 of Decin et al. (2018)\cite{Decin2018}. They are particularly visible in the CO, SiO and SO maps, both in the blue-shifted and red-shifted hemispheres, but more clearly in the former than in the latter. The interpretation of the high Doppler velocity components as streams rather than blobs is justified from their continuity with the slow wind: they only appear as blobs when considering a slice confined to an interval of Doppler velocity. Figure \ref{fig1} shows that they clearly stand out from the region of phase-space covered by the slow wind, making the distinction between high Doppler velocity wings and slow wind meaningful. This is further illustrated in Figure \ref{fig3}, which displays channel maps of the large Doppler velocity components for SO emission; other lines show similar patterns. We note the presence of an enhancement of emissivity near the line of sight in the lower $|V_z|$ intervals of the blue-shifted hemisphere, suggesting the presence of another component closer to the region of the slow wind. While the global picture is dominated by the former, which extends down to $-$20 \kms\ with intensity comparable to the red-shifted component, the presence of this enhancement cannot be ignored and needs to be studied in relation with the complex morpho-kinematics of the slow wind \cite{Nhung2019b}. However, this is beyond the scope of the present work that focuses on the dominant large Doppler velocity components.

\begin{figure}
  \centering
   \includegraphics[width=0.35\textwidth,trim=1cm 1cm 2.cm 2cm,clip]{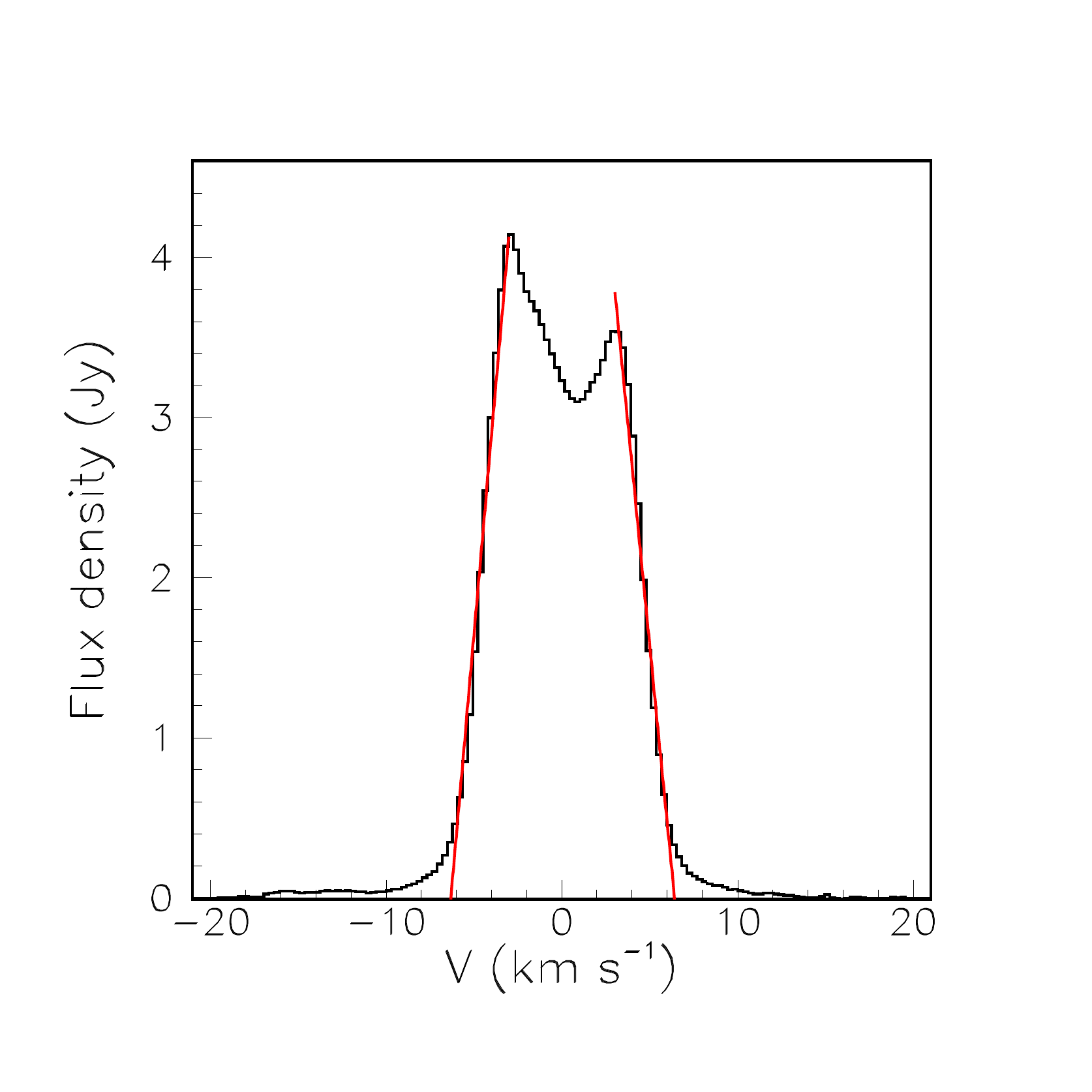}
  \caption{Doppler velocity spectrum of SO emission summed in the sky plane over a 1 arcsec radius circle centred on the star. The red lines are linear fits to the edge of the profile used to define the separation of the large Doppler velocity components from the slow wind.}
  \label{fig4}
\end{figure}

Globally, as illustrated in Figure \ref{fig4} for the case of SO emission, the $V_z$ spectra are dominated by the slow wind, the high velocity wings having much lower intensities on both the red-shifted and blue-shifted sides. This allows for a reliable evaluation of the end points of the Doppler velocity spectra of the slow wind, which we measure at $\sim$7.0, $\sim$8.3, $\sim$6.2, $\sim$6.0 and $\sim$6.8 \kms\ for CO, SiO, SO, SO$_2$ and HCN respectively on the red-shifted side. Very similar values are obtained on the blue-shifted side. As the maximal recoverable scales pertinent to each line are similar, these differences probably reveal different radial dependence of the molecular relative abundance and/or emissivity being probed along the line of sight. In particular, evidence for stream-like morphology is barely significant in the case of HCN as had first been noted by Decin et al. (2018)\cite{Decin2018}. The fact that the CO, SiO and SO$_2$ observations were made on a same day with a same antenna pattern while HCN was observed the day before with a different antenna pattern but a similar maximal recoverable scale is unlikely to explain the difference. We note that HCN is not expected to form in O-rich environments, its relatively strong emission in the slow wind may be explained by pulsation-induced shock-chemistry and the absence of detection at large Doppler velocities may simply be the result of insufficient sensitivity.

\begin{figure*}
  \centering
  \includegraphics[width=0.24\textwidth,trim=0.cm -1.cm 0.5cm 0.5cm,clip]{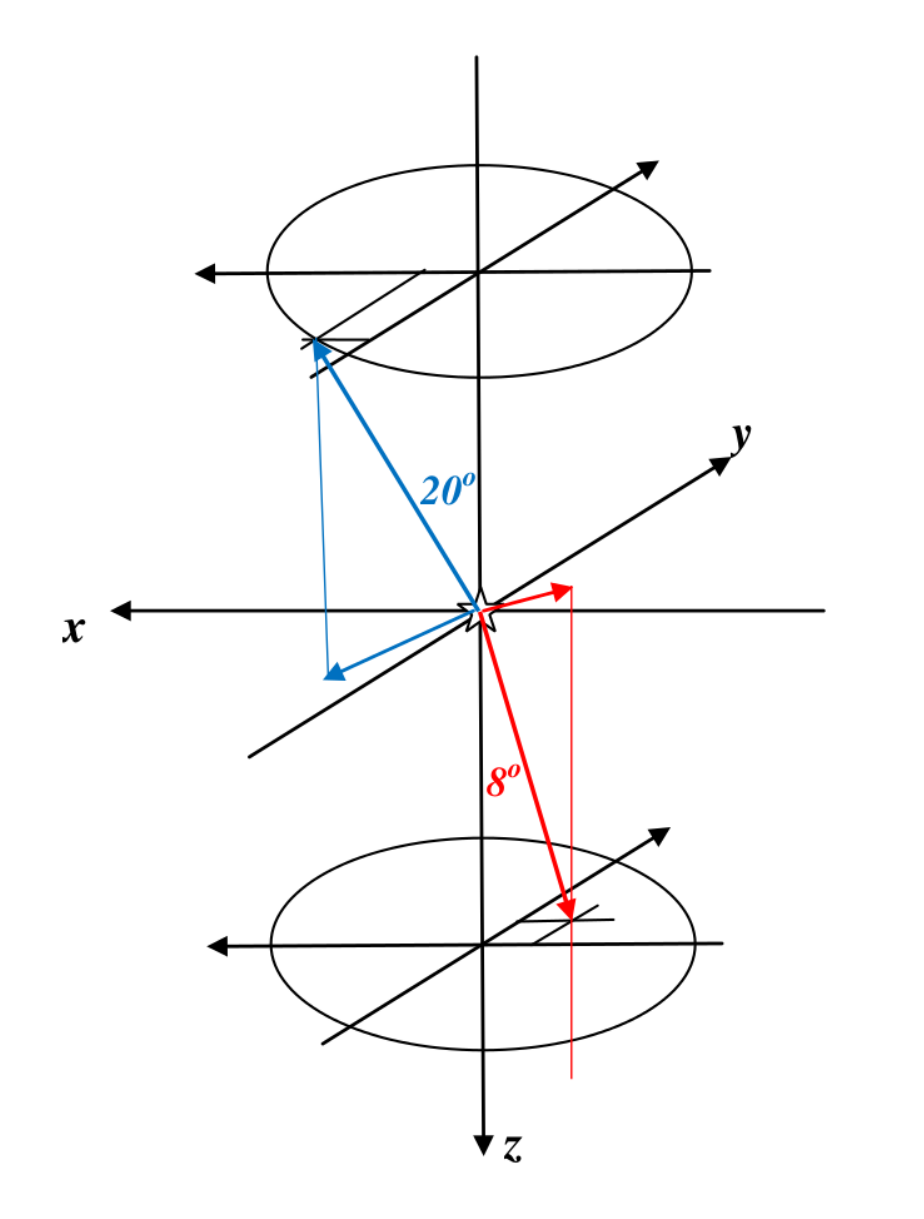}
  \includegraphics[width=0.6\textwidth,trim=0.cm 1.5cm 0.cm 1.5cm,clip]{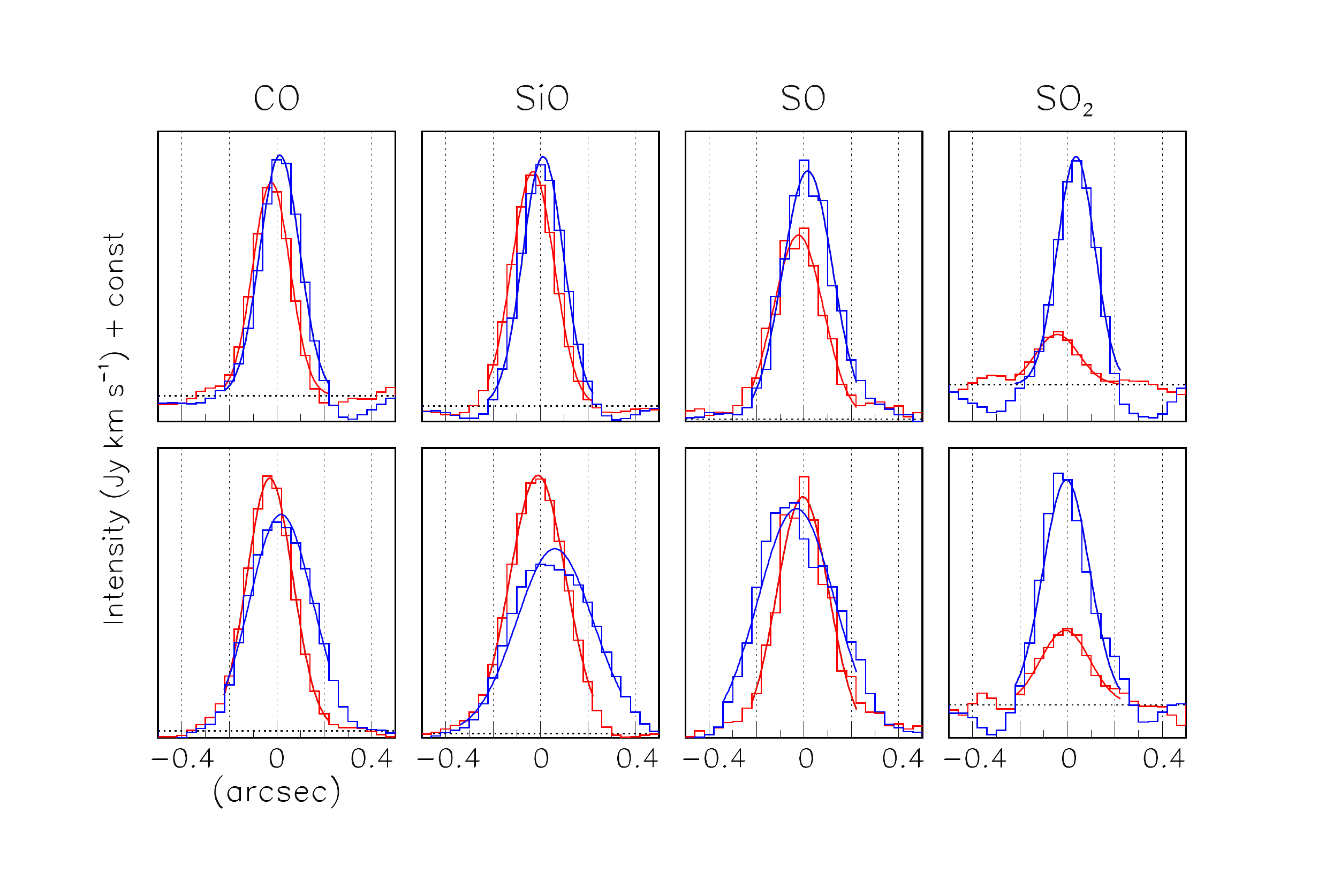}

  \caption{Left panel: illustration of a possible geometry assuming that the streams reach a velocity of 20 \kms\ over a distance of 60 au along the line of sight (see text). Right panels: dependence of the integrated intensity on $x$ (upper panels) and $y$ (lower panels) measured with respect to the stream axes defined by black arrows in Figure \ref{fig1}. Lines are labelled on top of the upper panels. The integration is made over  $|V_z|>$ 7.0, 8.3, 6.2 and 6.0 \kms\ for CO, SiO, SO and SO$_2$ respectively. For convenience, different arbitrary scales are used for different lines. Blue and red profiles are for blue-shifted and red-shifted hemispheres respectively. The curves show Gaussian fits.}
  
  \label{fig5}
\end{figure*}

The fact that the two candidate streams have the same reach in $|V_z|$ suggests that they are essentially symmetric with respect to the star. However, the red-shifted stream projects on the star within a beam size while the blue-shifted stream spans a significant $y$ interval, implying that such symmetry is not perfect. This may be because the streams are unrelated and that the same reach in $|V_z|$ is accidental. In this case, the streams may make very different angles $i_1$ and $i_2$ with respect to the line of sight as long as their velocities $V_1$ and $V_2$ obey $V_1$cos$i_1$=$V_2$cos$i_2$. But if the same reach in $|V_z|$ is not accidental the two streams are expected to be at small angle to each other and therefore at similar angles $i_1\sim i_2\sim i$ from the line of sight.  The effect of de-projection is essentially to change the $z$ scale to the extent that, to first order, $z$ may be approximated by a linear function of $V_z$. In particular, in principle, the large Doppler velocity components may be confined to the very close environment of the star. In order to get some idea of a possible geometry, we use an example illustrated in the left panel of Figure \ref{fig5}, where we assume arbitrarily that a stream velocity of 20 \kms\ is reached over a distance of 60 au along the line of sight. Approximating the stream projections on the P-V maps of Figure \ref{fig1} as shown by black arrows, we find that the red-shifted stream reaches this distance at $\sim -0.1$ arcsec in $x$ and $\sim$0.1 arcsec in $y$ while the blue-shifted stream reaches it at $\sim$0.1 arcsec in $x$ and $\sim-0.35$ arcsec in $y$. This means an angle of $\sim$8\dego\ between the red-shifted stream and the line of sight, an angle of $\sim$20\dego\ between the blue-shifted stream and the line of sight, and an angle of $\sim$14\dego\ between the two streams. Of course, these numbers scale with the arbitrary length of 60 au used in the example and are simply meant to illustrate a possible stream geometry.

The $x$ and $y$ profiles of the gas streams (excluding HCN) are illustrated in the right panels of Figure \ref{fig5}. They are referred to the stream axes indicated as black arrows in Figure \ref{fig1} and have Doppler velocities in excess of the slow wind end-point velocities listed above. On average, they are well centred to within $\pm$30 mas (1.8 au); Gaussian fits give standard deviations with respect to the mean of 90 mas in $x$ and 120 mas in $y$, meaning, after beam de-convolution, 70 mas (4.2 au) in $x$ and 90 mas (5.4 au) in $y$. The opening angle of the streams depends on their longitudinal extension; using as an example a mean distance of 30 au along the line of sight, this would correspond to an opening angle (standard deviation) of $\pm$9\dego.

While beyond the scope of the present article, we note the presence of significant depletions in well-defined regions of the data-cubes, in particular between Doppler velocities of $-1$ and 3 \kms. This is reminiscent of the blue-western depletion observed in EP Aqr by Tuan-Anh et al. (2019)\cite{TuanAnh2019}, who argue that it may be related to the nascent streams. In the present case, the complexity of the observed morpho-kinematics \cite{Nhung2019b} prevents from asserting reliably the existence of such a relation.

\subsection{Eliminating a possible effect related to continuum subtraction}\label{sec3.2}

The confinement of the red-shifted stream in the vicinity of the line of sight crossing the star in its centre comes as a surprise: at variance with EP Aqr, where axi-symmetry about this line of sight is well established, R Dor displays no obvious axi-symmetry and this line of sight does not seem to play any particular role in the complex morpho-kinematics of the circumstellar envelope \cite{Homan2018b, Vlemmings2018, Nhung2019b}.  In particular, the rotation axis observed by Homan et al. [12] makes an angle of only $20\pm20$\dego\ with the plane of the sky. This remark may suggest that the gas stream appearance of the high velocity wings is not real but is mimicked by some effect that has been overlooked. If such were the case, it would also shed doubts on the validity of the gas stream interpretation in the case of EP Aqr. It is therefore essential to review critically such possible effects. The next section discusses possible physical interpretations. Here, we address instead a possible effect of inadequate continuum subtraction.

\begin{figure}
  \centering
  \includegraphics[height=4.5cm,trim=0.5cm 1.cm 2.cm 2.cm,clip]{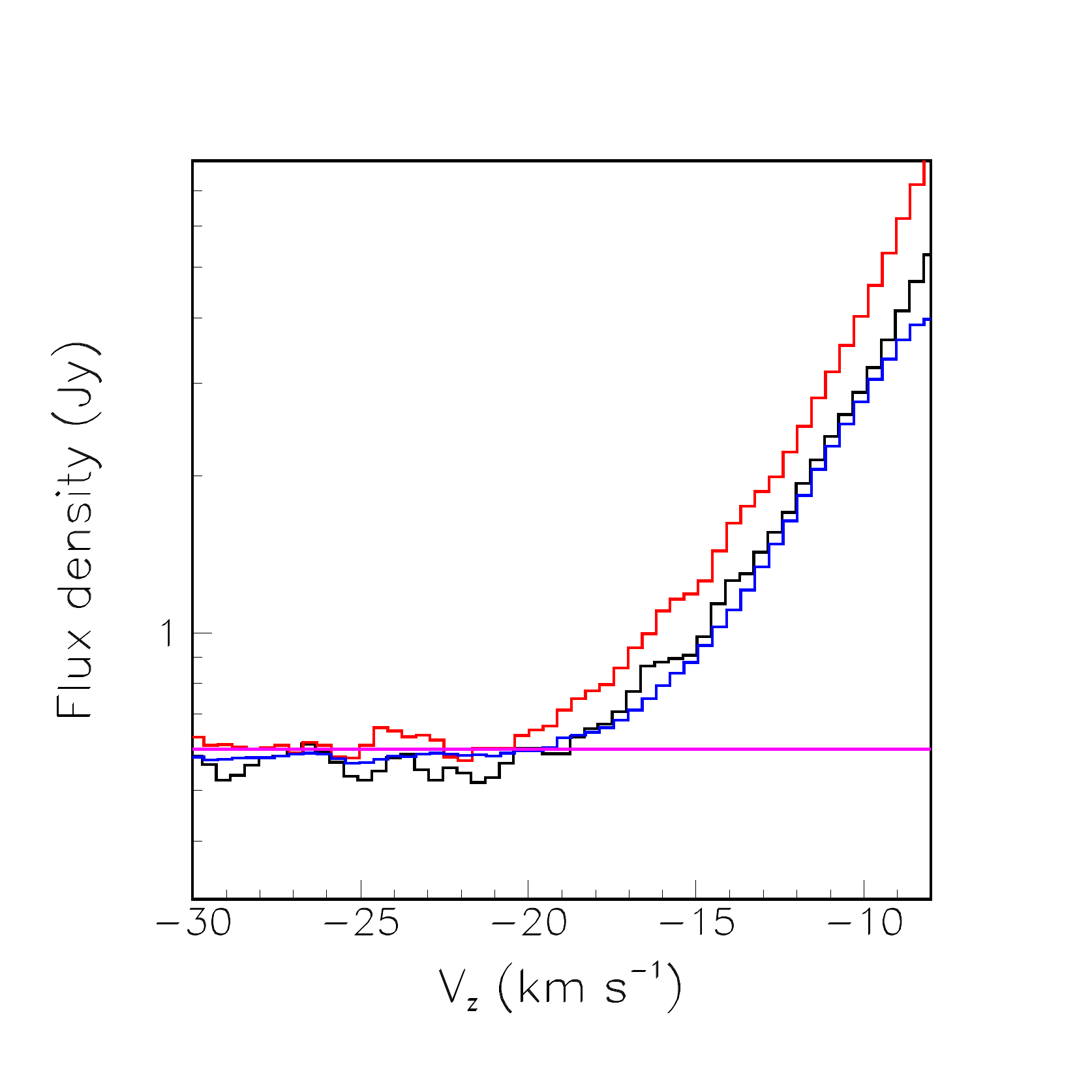}
  \includegraphics[height=4.5cm,trim=2.9cm 1.cm 2.cm 2.cm,clip]{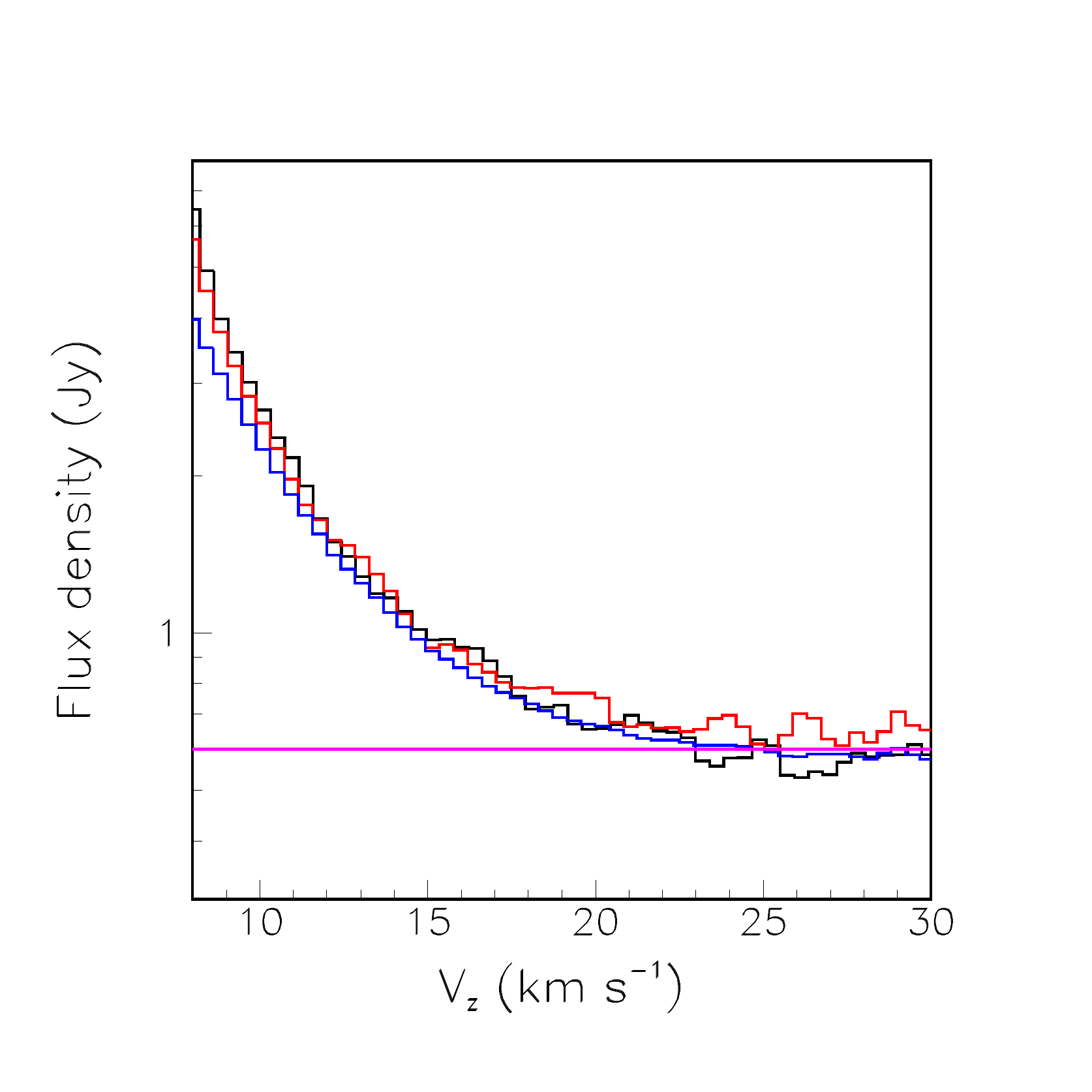}
  
  \caption{Doppler velocity distributions obtained for $|V_z|>8$ \kms\ using two datasets of SiO line emission having significantly different \textit{uv} coverage. The continuum, which has not been subtracted, is seen at the level of $0.6$ Jy. The dataset having the larger maximal recoverable scale is shown for $R<1.5$ arcsec (red) and for $R<0.3$ arcsec (blue). The dataset having the smaller maximal recoverable scale is shown for $R<1.5$ arcsec (black). }
  \label{fig6}
\end{figure}

As continuum emission is confined near the star and covers uniformly the observed range of Doppler velocities, inadequate continuum subtraction is a candidate for producing artefacts mimicking high velocity wings emitted along the line of sight near the origin of coordinates. The data used in the present work have been obtained after subtraction in the $uv$ plane of the contribution of continuum emission. An imperfection in the procedure of Fourier-transforming from the $uv$ plane to the sky plane or of producing the clean maps might have generated the undesired artefacts and have remained unnoticed. In order to check on this, we repeated the analysis of the SiO observations without performing any continuum subtraction, namely Fourier-transforming directly the measured visibilities. Not subtracting the continuum is a simplification and obviously does not affect significantly the main results in terms of beam size, maximum recoverable scale, etc. The result is illustrated in Figure \ref{fig6} using two datasets having different antenna configurations. The Doppler velocity distributions obtained for $|V_z|>8$ \kms\ display uniform continuum emission beyond $|V_z|>20$ \kms\ at a level of $\sim 0.60\pm0.03$ Jy on both red-shifted and blue-shifted sides. Here the uncertainty accounts for both noise levels and differences between different data sets. In comparison, the continuum level measured by Decin et al. [5] is 0.65 Jy with an uncertainty that does not exceed 0.01 Jy: our result is therefore $\sim 0.05\pm0.03$ Jy smaller. While of little relevance to the argument of the present article, this small difference is significant and probably reveals systematic differences in the data samples used and in their reduction. This result gives confidence in the reality of the high Doppler velocity components, seen to rise above continuum emission below $|V_z|\sim20$ \kms. It displays a remarkable symmetry between blue-shifted and red-shifted wings, at variance with an earlier interpretation of the ``blue blob'' as causing one-sided ejection of material \cite{Vlemmings2018}. It provides evidence for the emission to be essentially contained within an aperture radius of 0.3 arcsec (18 au). The mean projected acceleration, averaged over the ($V_z$ vs $y$) P-V maps of Figure \ref{fig1} is 0.68$\pm$0.10 \kms au$^{-1}$. However, to the extent that the large Doppler velocities are interpreted in terms of radial expansion, namely that blue- and red-shifted components are emitted from opposite sides of the star, the nearly perfect blue-red symmetry would suggest that effects of absorption and radiative transfer are small and can be ignored. 

\subsection{Interpretations in terms of pulsations or rotations}\label{sec3.3}

A gas stream interpretation implies that the streams are accelerated to nearly 20 \kms\ over long enough a distance to justify such a description. Quantitatively, we may get an idea of how long if we accept that the streams are approximately symmetric with respect to the star. The projected acceleration being estimated to reach $\sim$0.7 \kms au$^{-1}$ on the blue-shifted side, the Doppler velocity of 20 \kms\ is reached near 30 au projected distance from the star. If the blue-shifted stream were in the plane of the sky, it would reach a space velocity of 20 \kms\ in $\sim$ 30 au and so would the red-shifted stream.  But in projected distance the red-shifted stream stays within some 10 au from the star. Then, it would make an angle of $\cos^{-1}(10/30)=70$\dego\ with the plane of the sky (which contains the blue-shifted stream) contradicting the hypothesis that the two streams are nearly symmetric.  If the blue-shifted stream makes an angle $i_B$ and the red-shifted stream an angle $i_R$ with the line of sight, the common stream length is $30/\sin i_B=10/\sin i_R$ au, the right hand side of the equation being an upper limit , namely $\sin i_R<\nicefrac{1}{3}\,\sin i_B$. For $i_B=30$\dego\, $i_R<10$\dego\ corresponding to a stream length of 60 au. This very crude estimate gives the scale of acceptable stream lengths in the hypothesis of approximately symmetric streams; it corresponds to a space acceleration of $\sim$1.5 \kms\ au$^{-1}$.  If the acceleration was more sudden other interpretations, such as stellar pulsations or rotation, would become possible.

In the case of rotation, one would typically expect a line profile covering twice the maximal rotation velocity. Homan et al. (2018b)\cite{Homan2018b} quote a rotation velocity of only 12 \kms\ at a distance of 6 au from the star. Keplerian rotation would imply that the maximal observed velocity of 20 \kms\ would be reached at only $\sim$2.2 au from the star centre, very close to the stellar surface. The absence of detection in the $^{28}$SiO($\nu=1, J=8-7$) data analysed by  Homan et al. (2018b)\cite{Homan2018b} would then have to be blamed on insufficient sensitivity if a Keplerian rotation scenario was retained. Vlemmings et al. (2018)\cite{Vlemmings2018} quote a solid body rotation of 1.0$\pm$0.1 \kms\ at the stellar surface ($\sim$1.9 au) meaning that a rotation velocity of 20 \kms\ would be reached at a distance of $\sim$38 au from the star, nearly twice as far as the canonical limit that has been set. Unfortunately, the authors do not comment on the high Doppler velocity wings in general. They only illustrate the blue-shifted stream, which they claim to be one-sided, with a P-V diagram of $^{29}$SiO($\nu=0, J=5-4$) data that is consistent with the gas stream interpretation.

An interpretation in terms of a pulsating spherical shell is equally challenging. It has been explicitly excluded by Decin et al. (2018)\cite{Decin2018} on the basis of a model \cite{Nowotny2010}. For an expanding spherical shell to reproduce the narrow high velocity features observed in Figure 1, its radius must not exceed some 15 to 20 au. In principle, the very high angular resolution observations analysed by Vlemmings et al. (2018)\cite{Vlemmings2018} should allow for a measurement of the maximal Doppler velocity reached within an aperture at au scale. Unfortunately the authors do not address the issue: no measurement of the maximal velocity reached in a very small aperture is quoted.

Both interpretations, rotation and pulsation, predict red-blue symmetry  in contrast with what is observed. The possible presence of a companion, interfering with the blue-shifted high velocity wing, could then be invoked. Decin et al. (2018)\cite{Decin2018}, followed by Homan et al. (2018b)\cite{Homan2018b} and Vlemmings et al. (2018)\cite{Vlemmings2018} have noted that such presence would provide a welcome explanation of the large angular momentum implied by the observed rotation.  

\section{CONCLUSION}

Evidence has been obtained for a possible description of the nascent wind of AGB star R Dor in terms of a pair of high velocity gas streams emitted nearly back-to-back near the line of sight. They are reminiscent of those observed in similar conditions in the nascent wind of EP Aqr, an AGB star having properties very similar to R Dor. However, at variance with EP Aqr, the R Dor streams make a large angle with the rotation axes favoured by Vlemmings et al. (2018)\cite{Vlemmings2018} for the star and by Homan et al. (2018b)\cite{Homan2018b} for the rotating disc surrounding it. This may suggest that they be artefacts of improper data reduction and/or continuum subtraction but an analysis of continuum-unsubstracted data confirms the validity of the gas stream interpretation. Moreover, interpretations in terms of rotation or pulsations have been shown to meet important difficulties that prevent retaining such interpretations with reasonable confidence. A possibility that has not been seriously explored would be that the large Doppler velocities, rather than revealing large radial wind velocities, would instead reveal large line widths caused by turbulence, which would then be dominantly detected on the Earth side of the star atmosphere and would provide a simple explanation for the nearly perfect blue-red symmetry displayed in Figure 6. In principle, important relevant information should be contained in the very high angular resolution ALMA observations analysed by Vlemmings et al. (2018)\cite{Vlemmings2018} but, as far as we know, the issue has not been addressed by the authors in published material. The presence of large velocity components, reaching three times the terminal wind velocity, less than 100 au from each of two similar AGB stars, cannot be ignored when attempting a description of the mechanism governing the launch of their nascent wind and the breaking of the spherical symmetry present in the Red Giant phase. However, today, no fully convincing interpretation of the physics governing their existence can be reliably proposed. Future high resolution studies of other AGB stars displaying significantly higher Doppler velocities than the terminal wind will help clarifying this important issue.

\section*{ACKNOWLEDGEMENT}
We are grateful to Dr Ward Homan for sharing with us information on the analysis of R Doradus observations and to Professor Albert Zijlstra for support and advice. We thank the referees Professors Jan Martin Winters and Pierre Lesaffre for useful comments that helped improving the quality of the manuscript. This paper makes use of the following ALMA data: ADS/JAO.ALMA\#2017.1.00824.S and ADS/JAO.ALMA\#2013.1.00166.S. ALMA is a partnership of ESO (representing its member states), NSF (USA) and NINS (Japan), together with NRC (Canada) , MOST and ASIAA (Taiwan), and KASI (Republic of Korea), in cooperation with the Republic of Chile. The Joint ALMA Observatory is operated by ESO, AUI/NRAO and NAOJ. The data are retrieved from the JVO portal (http://jvo.nao.ac.jp/portal) operated by the NAOJ. We are deeply indebted to the ALMA partnership, whose open access policy means invaluable support and encouragement for Vietnamese astrophysics. Financial support from the World Laboratory, VNSC and NAFOSTED is gratefully acknowledged. This research is funded by the Vietnam National Foundation for Science and Technology Development (NAFOSTED) under grant number 103.99-2018.325. 

\bibliographystyle{ciprefstyle-unsrt}

\end{document}